\documentclass[twocolumn,showpacs,preprintnumbers,amsmath,amssymb]{revtex4}

\usepackage{epsfig}

\begin{document}

\title{DIS structure functions and the double-spin asymmetry\\
in  $\rho^0$ electroproduction within a Regge approach}

\author{N.I. Kochelev}
 \email{kochelev@thsun1.jinr.ru}
 \affiliation {Bogoliubov Laboratory of Theoretical Physics,
     JINR, Dubna, Moscow region, 141980 Russia}
 \altaffiliation[Also at ]{Institute of Physics and Technology, Almaty,
480082, Kazakhstan}
\author{K. Lipka}
 \email{Katerina.Lipka@desy.de}
\affiliation{DESY Zeuthen, 15738 Zeuthen, Germany}
\author{W.-D.~Nowak}
 \email{Wolf.Dieter.Nowak@desy.de}
\affiliation{DESY Zeuthen, 15738 Zeuthen, Germany}
\author{V. Vento}
 \email{vicente.vento@uv.es}
 \affiliation{Departament de F\'{\i}sica Te\`orica and Institut de F\'{\i}sica
     Corpuscular, Universitat de Val\`encia-CSIC E-46100,
     Burjassot (Valencia), Spain}
\author{A.V. Vinnikov}
 \email{vinnikov@thsun1.jinr.ru}
\affiliation{Bogoliubov Laboratory of Theoretical Physics,
     JINR, Dubna, Moscow region, 141980~Russia}
 \altaffiliation[Also at ]{School of Physics,
Seoul National University, Seoul 151-747, Korea}

\date{\today}

\begin{abstract}
The proton, neutron and deuteron structure functions $F_2(x,Q^2)$ and
$g_1(x,Q^2)$, measured at intermediate $Q^2$, are analyzed
within a Regge approach. This analysis serves to fix the parameters
of this scheme which are then used to calculate, in a unified Regge approach,
the properties of $\rho^0$ meson electroproduction on the proton and the
deuteron. In this way, the double-spin asymmetry observed at  HERMES in
$\rho^0$ electroproduction on the proton, can be related to the
anomalous behavior of the flavor-singlet part of the spin-dependent
structure function $g_1(x,Q^2)$ at small $x$.
\end{abstract}


\maketitle

\section{Introduction}
Understanding the spin structure of the nucleon is today one of the main
problems in hadron physics. Polarized beams have allowed significant progress
in the measurement of the nucleon spin-dependent structure functions
in the Deep Inelastic Scattering (DIS) regime.
A result of some of these measurements~\cite{g1dat} is that, at small values
of Bj{\o}rken $x$, the neutron spin-dependent  structure function $g_1^n(x,Q^2)$,
which according to the SU(6)$_W$ model is dominated by the flavor singlet
quark sea part, exhibits an anomalous behavior.
More precisely, at $x\to~0$, $g_1^n(x,Q^2)\sim 1/x$, while the
expected behavior should be $1/x^{\alpha_{a_1}}$, where $a_1$ is
a conventional
Regge trajectory with appropriate
quantum numbers and intercept $\alpha_{a_1}=-0.5\div 0.5$.
Experimental data are thus well
described in terms of a leading trajectory with an intercept close to 1,
a value
reminiscent of the Pomeron. In fact, this type of behavior at small
$x$ was predicted by leading order perturbative QCD
calculations~\cite{rysking1}.
However, since next-to-leading order corrections to
the perturbative BFKL Pomeron~\cite{bfkl} are large~\cite{lipatov},
it is not clear to which extent the perturbative description is valid.

In Ref.~\cite{our1}, the observed anomalous behavior
of $g_1^n(x,Q^2)$, was assumed to signal the existence of a flavor-singlet
exchange with large intercept
and negative signature, which was then used to explain the experimental
data on diffractive hadronic reactions: vector meson photoproduction at
$\sqrt{s}\approx$ 100 GeV, $|t|$=1$\div$10 GeV$^2$ and
proton-proton elastic scattering at $\sqrt{s}$=20$\div$ 60 GeV,
$|t|$=2$\div$10 GeV$^2$~\cite{our1,our11}. This exchange gives sizeable contributions
to the differential cross-sections at large energy due to the large value of
its intercept. We called it $f_1$ since its quantum numbers coincide with those
of the axial vector $f_1(1285)$ meson (P=C=+1, negative signature).
The mechanism leading to this exchange in QCD is not known. The relation
of the Pomeron with the scale anomaly, led us to conjecture  a relation
between the $f_1$ and the gluonic axial anomaly, since the $f_1$ strongly
mixes with gluonic degrees of freedom~\cite{our1}.

In order to investigate this unnatural-parity exchange the best observables
are spin-dependent cross-sections. For example, we already have shown that the
$f_1$ exchange leads to large double-spin asymmetries in different diffractive
reactions~\cite{our1,our2,our3}.

For $s$-channel helicity conserving (SCHC) reactions it seems reasonable,
in the context of the VMD model, to establish a relation between the matrix
elements describing the structure functions and vector meson electroproduction
at low $-t$, as indicated in Fig.~1. This relation arises naturally in the
context of
the Regge approach~\cite{collins}, where the imaginary part of the forward
Compton scattering amplitude, relevant to describe DIS structure functions,
and the total amplitude for vector meson production at $-t\rightarrow 0$ are
connected.

The Regge model is a well known approach to describe diffractive
reactions with small momentum transfer and high energy, a kinematic region
where non-perturbative QCD effects are important.
For this reason the  investigation of diffraction  and structure functions
at low $x$  are important items included into the experimental programs
of the current and future accelerators: HERA, TEVATRON,  RHIC, LHC, etc.
These studies are expected to considerably enlarge the insight into the nature
of strong interactions between quarks at large distances~\cite{levin}.
Several scenarios for the direct relation between the complex structure of the
QCD vacuum and Regge behavior have been suggested
\cite{simonov,kharzeev,shuryak}.

The effective propagator for the spin-nonflip amplitude for scattering
of two particles with momenta $p_1$ and $p_2$ has the following form in
Regge theory:
\begin{eqnarray}
D_R(s,t)=-\frac{1}
{sin\bigl (\pi\alpha_R(t) \bigr )\Gamma\bigr ( \alpha_R(t)\bigl ) }
\Bigl ( \frac{s}{s_0} \Bigr ) ^{\alpha_R(t)-1}\times \nonumber \\
\times \Bigl [ 1+\sigma_R cos\bigl (\pi\alpha_R(t)\bigr ) -
\imath\cdot\sigma_R sin\bigl ( \pi\alpha_R(t)\bigr ) \Bigr ] , \label{prop}
\end{eqnarray}
where $s=(p_1+p_2)^2$, $t=(p_1-p_2)^2$, $s_0$ is
$max(p_1^2,p_2^2,p_1'^2,p_2'^2)$, $\alpha_R(t)$ describes the Regge trajectory
of the particles exchanged in the $t$ channel and $\sigma_R$
their signature, related to the spin $J$ by $\sigma=(-1)^J$. The $\Gamma$
function in the denominator provides the correct analytic behavior of the
amplitude~\cite{laget97}. In many cases, a simpler
formula can be used~\cite{manaenk}, but we prefer the original
form in order to keep a precise relation between the real and the imaginary
parts of the amplitude. This relation will be used below to establish
a connection between the properties of the DIS structure functions and
those of vector meson electroproduction off the nucleon as suggested above.

Reggeons, with signature and space parities of the same sign, $\sigma=P$
(natural parity), have an amplitude which does not depend on the helicities
of the colliding particles. Their exchange gives the dominant contributions
to the spin-independent nucleon structure functions and to the  total
cross-sections.
Exchanges with unnatural parity, $\sigma=-P$, have an  amplitude which is
proportional to the product of the particle helicities. Thus, only Reggeons
with unnatural parities can contribute to the spin-dependent structure
function $g_1(x,Q^2)$. Furthermore, a longitudinal double-spin asymmetry
in  two-particle scattering only arises when an unnatural-parity-exchange
contribution to the scattering amplitude exists.

The main goal of this paper is an analysis of the deep-inelastic structure
functions $F_2(x,Q^2)$ and $g_1(x,Q^2)$ in the framework of the Regge model;
the extracted Reggeon parameters will be used to calculate the elastic
cross-section, and finally the double-spin asymmetry, of $\rho^0$
electroproduction for intermediate values of $Q^2$.

The Regge scheme will be defined by the exchange of trajectories with very
large intercepts (Pomeron and the `anomalous' component of $f_1$) and of
secondary Regge trajectories, namely $f_2$, $a_2$, the `normal' component of
$f_1$ (which will be called $f_1^\prime$) and  $a_1$. Especially the
contributions of secondary Reggeons are considered to be of importance
at HERMES energies, $\sqrt{s}=7$ GeV.

\section{The structure function $F_2(x,Q^2)$ in a Regge approach
at intermediate values of $Q^2$}
The parameters of the natural-parity Reggeons in soft interactions
($Q^2<$ 1 GeV$^2$) are quite well known. They can be obtained from fits
to the total hadronic cross-sections~\cite{landtot}. In particular,
the parameters of the Pomeron, $f_2$ and $a_2$ Reggeon trajectories
which are relevant for our calculation, are given by the Particle Data
Group~\cite{pdg}:
\begin{eqnarray}
\alpha_P(t)&=&1.093+0.25\;t,\beta_{Pqq}=\beta_{PNN}/3=1.6~{\rm GeV}^{-1},
\nonumber \\
\alpha_{f_2}(t)&=&0.358+0.9\;t,\beta_{f_2 qq}=\beta_{f_2 NN}/3=
4.7~{\rm GeV}^{-1},
\nonumber\\
\alpha_{a_2}(t)&=&0.545+0.9\;t,\beta_{a_2 qq}=\beta_{a_2 NN}=
4.9~{\rm GeV}^{-1},\label{par}
\end{eqnarray}
where $\beta_{RNN}$ and $\beta_{Rqq}$ are the coupling constants for the
Reggeon-nucleon and Reggeon-quark couplings, respectively. The effective
Lorentz structure of the coupling of these Reggeons to nucleons and quarks
is assumed to be $\gamma_{\mu}$~\cite{landnacht}. The values of the slopes
of the trajectories, which cannot be obtained from a fit to the total
hadronic cross-sections, are taken from Ref.~\cite{collins}. The relation
between the couplings to quarks and nucleons is obtained from the
naive constituent quark model.
A clear separation of the secondary Reggeons and the Pomeron is possible
because the energy dependencies of their amplitudes described by
the intercepts $\alpha(0)$, $\sigma^{tot}_R\sim s^{\alpha_R(0)-1}$
are very different. Specifically, the Pomeron contribution to the total
hadronic cross-section rises as $s^{0.093}$, while, for example, the $f_2$
contribution falls as $s^{-0.642}$.

For large virtualities of the scattering particles, the effective intercepts
of  the Pomeron and  the Reggeons change~\cite{kaidalov}.  In particular,
it was measured by the H1 and ZEUS Collaborations at HERA~\cite{H1,ZEUS}
that the effective Pomeron intercept, extracted from the analysis
of the $F_2(x,Q^2)$ structure function, ranges
from about $1.1$ at
$Q^2\approx 0$ to about  $ 1.4$ at $Q^2>10^2$ GeV$^2$. Therefore, to make
predictions for
HERMES ($<Q^2>$=1.7 GeV$^2$), knowledge of the Pomeron and Reggeons
intercepts at $Q^2$ values of a few GeV$^2$ is required.

The Regge formulae~\cite{allm} for the proton and deuteron structure
functions used are
\begin{eqnarray}
F_2^p(x,Q^2)&=&\frac{Q^2}{Q^2+\mu_N^2} \times\nonumber \\
&\times & \Bigl ( \frac{A_P}{x^{\alpha_P(0)-1}}+
\frac{A_{f_2}}{x^{\alpha_{f_2}(0)-1}}+\frac{A_{a_2}}
{x^{\alpha_{a_2}(0)-1}}\Bigr ) , \nonumber\\
F_2^d(x,Q^2)&=&\frac{Q^2}{Q^2+\mu_N^2}\Bigl ( \frac{A_P}{x^{\alpha_P(0)-1}}+
\frac{A_{f_2}}{x^{\alpha_{f_2}(0)-1}}\Bigr ) \label{f2struct}.
\end{eqnarray}
The various structure functions appearing in these equations represent the
$SU(3)_f$ flavor singlet (the Pomeron), the $SU(2)_f$ isoscalar ($f_2$) and
the isovector ($a_2$) exchanges. The parameter $\mu_N$ effectively accounts
for the scaling violation at small $Q^2$ for the natural parity exchanges.
Using Eq.~(\ref{f2struct}) we made a fit (not shown) to the recent
$F_2$  data~\cite{f2data} at
$Q^2=1\div 3$ GeV$^2$ and $x\lesssim0.1$ (Regge region).
 The extracted parameters
of Pomeron, $f_2$ and $a_2$ Reggeons are
\begin{eqnarray}
A_P=0.15\pm0.02,~~~\alpha_P(0)=1.20\pm0.01; \nonumber \\
A_{f_2}=0.46\pm0.05,~~~\alpha_{f_2}(0)=0.61\pm0.07; \nonumber \\
A_{a_2}=0.12\pm0.11,~~~\alpha_{a_2}(0)=0.36\pm0.28;\nonumber\\
\mu_N=0.80\pm0.02 \rm{~GeV}. \label{nonpol}
\end{eqnarray}
This fit provides a value of $\chi^2=191.2$ for 164 data points
($\chi^2/ndof$=1.17). It is evident that already in the semi-hard region
($Q^2$-values of a few GeV$^2$) modified Regge intercepts
have to be used. The slopes can be assumed to remain unchanged as they
seem to vary quite slowly in this $Q^2$ region~\cite{hermschc}.
Moreover, they have no strong effect on the double-spin asymmetry.

\section{The spin-dependent structure function $g_1(x,Q^2)$
in a Regge approach.}
Parameters for unnatural-parity Reggeons have been extracted earlier
using Regge-type {\it phenomenological fits} to the spin-dependent
structure function $g_1(x,Q^2)$~\cite{teryaev,bianchi}. In our approach
not only effective intercepts of Regge exchanges are determined, but
also their couplings  with nucleons and quarks. These couplings are
a necessary input to the calculation of the double-spin asymmetry in
vector meson electroproduction (see below). We use the relation
between $g_1(x,Q^2)$ and the imaginary part of the spin-dependent  forward
Compton scattering
amplitude. By {\it direct calculation} (Fig.~1a)  the contribution of the
unnatural-parity-Reggeon exchanges to  this amplitude is obtained as:
\begin{eqnarray}
T^{\mu\nu}(q,p,S)=\beta_{Rqq}\beta_{RNN}C_R
\bar u (p,S)\gamma_{\alpha}\gamma_5 u(p,S)
\times \nonumber \\
\times R^{\mu\nu\alpha} \frac{1- exp \{ -i\pi\alpha_R(0)\} }
{sin\{ \pi\alpha_R(0)\} \Gamma \bigl ( \alpha_R(0) \bigr ) }
\Bigl ( \frac{2p\cdot q}{s_0} \Bigr ) ^{\alpha_R(0)-1},
\label{rowg1}
\end{eqnarray}
where we assume that all exchanges $f_1$,  $ f_1^\prime $ and $a_1$
have a $\gamma_\alpha\gamma_5$
Lorenz structure for both the quark and the nucleon vertices. The
nucleon spin is denoted by $S$, and $s_0\approx Q^2$ is the characteristic
scale in the process.

The factor $C_R$ is given by $e_u^2+e_d^2+e_s^2$ for the $SU(3)_f$ singlet
`anomalous' exchange ($f_1$), by $e_u^2+e_d^2$ for the $SU(2)_f$ singlet
'normal' exchange ($f_1'$) and by $e_u^2-e_d^2$ for the $SU(2)_f$ triplet
$a_1$; $R^{\mu\nu\alpha}$ is given by Ref.~\cite{rosen}
\begin{eqnarray}
R^{\mu\nu\alpha}=-2\int \frac{d^4 k}{(2\pi)^4}
\frac{1}{(k^2-m_q^2)((k-q)^2-m_q^2)^2 } \times \\
Tr \bigl {[} (m_q+\hat k+\hat q)
\gamma^{\nu}(m_q+\hat k)\gamma^{\nu}(m_q+\hat k +\hat q)
\gamma^{\alpha}\gamma_5 \bigr {]}, \nonumber
\end{eqnarray}
where $m_q$ is the mass of the quark in the quark loop in Fig.~1.

Using the relation $\bar u (p,S)\gamma_{\alpha}\gamma_5 u(p,S)=2m S_{\alpha}$,
and the formula
\begin{equation}
R^{\mu\nu\alpha}=\frac{1}{2\pi^2}q_{\tau}\epsilon^{\tau\mu\nu\alpha},
\end{equation}
for $-q^2>> m_q^2$~\cite{efremov} we have
\begin{eqnarray}
T^{\mu\nu}(q,p,S)=\frac{m}{\pi^2}C_R
\beta_{Rqq}\beta_{RNN} S_{\alpha}q_{\tau}
\epsilon^{\tau\mu\nu\alpha} \times \\
\times\frac{1-exp \{ -i\pi\alpha_R(0)\} }
{sin\{ \pi\alpha_R(0)\} \Gamma \bigl ( \alpha_R(0) \bigr ) }
\Bigl ( \frac{2p\cdot q}{Q^2} \Bigr ) ^{\alpha_R(0)-1}. \nonumber
\end{eqnarray}
This expression is compared to the general form
\begin{eqnarray}
T^{\mu\nu}_{(spin)}&=&
\imath \epsilon^{\mu\nu\alpha\beta}q_{\alpha}S_{\beta}\frac{m}{pq}T_3+ \\
&+&\imath \epsilon^{\mu\nu\alpha\beta}q_{\alpha}(S_{\beta}(p\cdot q)
-p_{\beta}(S\cdot q) )\frac{m}{(p\cdot q)^2} T_4, \nonumber
\end{eqnarray}
where the DIS structure functions $g_1$ and $g_2$ are related to the
amplitudes $T_3$, $T_4$ by

\begin{equation}
g_1(q,\nu)=-\frac{Im(T_3)}{2\pi},~~~g_2(q,\nu)=-\frac{Im(T_4)}{2\pi},
\end{equation}
and where $\nu=p\cdot q$.

By assuming that the  structure function
$g_2$  is small, we obtain
\begin{equation}
g_1(q,\nu)=\frac{\nu C_R\beta_{Rqq}\beta_{RNN}}
{2\pi^3 \Gamma \bigl (\alpha_R(0)\bigr )}
\Bigl ( \frac{\nu}{Q^2}\Bigr )^{\alpha_R(0)-1}_{\cdot}
\end{equation}

Using $x=Q^2/2 \nu$ we have
\begin{equation}
g_1(x,Q^2)=\frac{C_R\beta_{Rqq}\beta_{RNN}}
{4\pi^3 \Gamma \bigl (\alpha_R(0)\bigr )}
\frac {Q^2}{x^{\alpha_R(0)}}.
\end{equation}
This form does not yet provide the correct scaling behavior for the
structure function at $Q^2\to \infty$ since the non-locality
of the  Reggeon-quark vertex in Fig.~1a was not taken into account.
Phenomenologically, this can be accomplished
by introducing a form factor (as has been done  in the Pomeron case~\cite{landrho}):
\begin{equation}
F(Q^2)=\frac{\mu_U^2}{Q^2+\mu_U^2},
\end{equation}
where $\mu_U$ is the effective parameter describing the scaling violation
at low $Q^2$ for the unnatural-parity Reggeons. Then the contribution of
a Reggeon to the $g_1(x,Q^2)$ structure function reads
\begin{equation}
g_1(x,Q^2)=\frac{\mu_U^2 C_R \beta_{Rqq}\beta_{RNN}}
{4\pi^3 \Gamma \bigl (\alpha_R(0)\bigr )}
\frac{Q^2}{Q^2+\mu_U^2}\frac{1}{x^{\alpha_R(0)}}
\label{g1}
\end{equation}
which provides the correct scaling behavior of the structure
function at large $Q^2$.

The parameters of the Reggeons with unnatural parity can be found
from a fit to the available data on polarized DIS. Since data exist for
proton, neutron and deuteron targets, this
 allows the separation of the
isoscalar and isovector parts of the respective  structure functions
\begin{eqnarray}
g_1^p(x,Q^2)&=&g_1^{(1)}(x,Q^2)+g_1^{(3)}(x,Q^2), \nonumber \\
g_1^n(x,Q^2)&=&g_1^{(1)}(x,Q^2)-g_1^{(3)}(x,Q^2),\nonumber \\
g_1^d(x,Q^2)&=&g_1^{(1)}(x,Q^2),
\label{sf}
\end{eqnarray}
where $g_1^{(1)}(x,Q^2)$ and $g_1^{(3)}(x,Q^2)$ are the isoscalar and
isovector components, respectively.

It was pointed out~\cite{teryaev} that the isoscalar part contains
two components: `normal' and `anomalous'. The `normal' component
has an intercept of $\alpha(0)\approx 0.5$ while the `anomalous'
has a large intercept, $\alpha(0)$ close to 1.
Using Eq.~(\ref{g1}), the structure
functions in Eq.~(\ref{sf})  can be written
\begin{eqnarray}
g_1^{(1)}(x,Q^2)&=&\frac{2}{3}\frac{\mu_U^2\beta_{f_1 qq}\beta_{f_1 NN}}
{4\pi^3  \Gamma \bigl (\alpha_{f_1}(0)\bigr ) }
\frac{Q^2}{Q^2+\mu_U^2}\frac{1}{x^{\alpha_{f_1}(0)}}+ \nonumber \\
&+&\frac{5}{9}\frac{\mu_U^2\beta_{f_1' qq}\beta_{f_1' NN}}
{4\pi^3  \Gamma \bigl (\alpha_{f_1'}(0)\bigr ) }
\frac{Q^2}{Q^2+\mu_U^2}\frac{1}{x^{\alpha_{f_1'}(0)}},\nonumber\\
g_1^{(3)}(x,Q^2)&=&\frac{1}{3}\frac{\mu_U^2\beta_{a_1 qq}\beta_{a_1 NN}}
{4\pi^3  \Gamma \bigl (\alpha_{a_1}(0)\bigr ) }
\frac{Q^2}{Q^2+\mu_U^2}\frac{1}{x^{\alpha_{a_1}(0)}}.
\end{eqnarray}
For  the fit we use the available data on $g_1(x,Q^2)$~\cite{g1dat} in
the same kinematic region ($Q^2=1\div 3$ GeV$^2$ and
 $x\lesssim0.1$)
which has been used above in fitting  $F_2(x,Q^2)$.
Unfortunately, the amount of data available from polarized
          lepton-nucleon scattering is much smaller than
          that in the unpolarized case. The number of fit parameters
          therefore had to be decreased in order to get a high fit quality.
 Thus we  assume a degeneracy of the
'normal' isovector $f_1^\prime $ and
isoscalar $A_1$ trajectories, analogous to the
well known case of the degeneracy of $\omega$ and $\rho$ trajectories.
Moreover  we take their coupling to quarks to be equal.
  Under these conditions
the  total contribution of the$f_1^\prime$ and $A_1$  to the neutron  structure
 function $g_1^n$ exactly vanishes,
as it should be in a valence $SU(6)_W$ model for nucleon structure
functions.

The resulting parameters are:
\begin{eqnarray}
\alpha_{f_1}(0)&=&0.88\pm 0.14,\label{anparr}\\
\beta_{f_1 qq}\beta_{f_1 NN}&=&-3.04\pm2.42 ~\rm{GeV}^{-2},
\nonumber\\
\alpha_{f_1'}(0)&=&\alpha_{A_1}(0)=0.62\pm0.13,\nonumber\\
\beta_{f_1' qq}\beta_{f_1' NN}&=&\frac{3}{5}
\beta_{A_1 qq}\beta_{A_1 NN}=13.57\pm9.79~\rm{GeV}^{-2};\nonumber \\
\mu_U&=&1.45\pm0.59~\rm{GeV},
\label{parameters}
\end{eqnarray}
which yield a common $\chi^2$ value of 31.8 for 53 data points (22 for $g_1^p$,
14 for $g_1^n$ and 17 for $g_1^d$). We found that using different
values of $\mu_U$ for the `anomalous', isoscalar and isovector components
does not provide a substantial improvement of the $\chi^2$ value.Therefore
we used the same $\mu_U$ for the three channels.

As before, we take that the slopes of the  $f_1^\prime$ and
$a_1$ exchanges to be the standard Reggeon value, i.e.
0.9 GeV$^{-2}$, while for the $f_1$ exchange we take the
slope to be zero, as  established in Ref.~\cite{our1}. We mention again
that the effect of the slopes on the double-spin asymmetry is negligible.

The result of our fit
to $g_1(x,Q^2)$ in the Regge region $x\lesssim 0.1$,
presented in Fig.~\ref{g1pp}-
\ref{g1dd},
  leads to the
conclusion that our model is in
 good agreement with the available data.
This supports the expectation that it may
describe not only the
structure functions $F_2$ and $g_1$, but -- in the context
of SCHC -- also vector meson production at small values of $-t$.

\section{The cross-section of vector meson electroproduction}

The cross-section of vector meson electroproduction reads (cf. Fig.~1b):
\begin{equation}
\frac{d\sigma}{d|t|}=
\frac{|M(t)|^2.}{64\pi W^2 ({\vec q}_{cm})^2},
\label{cross}
\end{equation}
where $({\vec q}_{cm})^2=\bigl ( W^4+Q^4+m_p^4+2W^2Q^2-2W^2m_p^2+2Q^2m_p^2
\bigr )/ 4W^2$, $W$ is the center-of-mass energy of the virtual-photon proton
system and $m_p$ is the proton mass.

It was found~\cite{our2,landrho,laget}
that the main features of the photoproduction reaction
can be reproduced within a simple
non-relativistic model for the vector meson wave function, where the quark
and the anti-quark
form the meson only if they have equal momenta. Above $Q^2\approx$ 3 GeV$^2$,
the quark's off-shellness and Fermi
motion inside the vector meson have to be taken into account~\cite{cudell}.
At smaller $Q^2$ these effects are not important
and the non-relativistic model is applicable. In this framework,
the Pomeron, $f_2$ and  $A_2$  Reggeon exchange amplitudes read~\cite{our2}

\begin{eqnarray}
M_N & = & 4C_R^N m_V \beta_{Rqq}\beta_{RNN}
\sqrt{\frac{3 m_V \Gamma_{e^+ e^-}}{\alpha_{em}}}
\bar u (p_2)\gamma_{\alpha} u(p_1) \times \nonumber\\
& \times &
\frac{(g_{\mu\nu} q^{\alpha} -g_{\nu\alpha} p_V^{\mu}-g_{\mu\alpha}q^{\nu})
\varepsilon_{\gamma}^{\mu} \varepsilon_V^{\nu}}
{q^2+t -m_V^2} \times \nonumber \\
&\times &\frac{1+exp \{ -i\pi\alpha(t)\} }
{sin\{ \pi\alpha(t)\} \Gamma \bigl (\alpha_R(t)\bigr ) }
\Bigl ( \frac{s}{Q^2} \Bigr ) ^{\alpha_R(t)-1}
F_R^V(t) \approx \nonumber \\
& \approx & \frac{8 m_V \beta_{Rqq}\beta_{RNN} Q^2}{Q^2-t +m_V^2}
\sqrt{\frac{3 m_V \Gamma_{e^+ e^-}}{\alpha_{em}}}\times \\
&\times & \frac{1+exp \{ -i\pi\alpha(t)\} }
{sin\{ \pi\alpha(t)\}  \Gamma \bigl (\alpha_R(t)\bigr ) }
\Bigl ( \frac{s}{Q^2} \Bigr ) ^{\alpha_R(t)}
F_R^V(t), \nonumber
\label{pomampl}
\end{eqnarray}
where $m_V$ is the mass of the vector meson,
$C_R^N=e_u-e_d$ for the Pomeron and  $f_2$
exchanges, while $C_{a_2}^N=e_u+e_d$,
 and  $\Gamma_{e^+ e^-}$ is
its leptonic width. The  parameters of the exchanges have been fixed
above :
$\beta_{Pqq}=\beta_{PNN}/3=1.6$ GeV$^{-1}$, $\beta_{f_2 qq}=\beta_{f_2 NN}/3=
4.7$ GeV$^{-1}$,$\beta_{a_2 qq}=\beta_{a_2 NN}=
4.9~${\rm GeV}$^{-1}$,
 $\alpha_P(t)=1.20+0.25\;t$, $\alpha_{f_2}(t)=0.64+0.9\;t$.
$\alpha_{a_2}(t)=0.36+0.9\;t$.
The vector form factor of the Pomeron-NN vertex is~\cite{landrho}
\begin{equation}
F_P^V=\frac{4m_p^2-2.8\;t}{(4m_p^2-t)(1-t/0.71)^2}.
\end{equation}
We assume that the Pomeron and $f_2$ Reggeon vertices have the same
form factors. An additional factor has to be included to take into
account the non-locality of these vertices,
\begin{equation}
\frac{\mu_N^2}{\mu_N^2+Q^2-t},
\end{equation}
where $\mu_N$ = 0.8 GeV according to Eq.~(\ref{nonpol}).

In a similar way one can obtain the unnatural-parity
Reggeon contribution to the vector meson production amplitude~\cite{our2}.
\begin{eqnarray}
M_U = 4 m_V C_R^U
 \beta_{Rqq}\beta_{RNN}
\sqrt{\frac{3 m_V \Gamma_{e^+ e^-}}{\alpha_{em}}}\times \nonumber \\
\times
\frac{\epsilon_{\mu\nu\alpha\beta}q^{\beta} \varepsilon_{\gamma}^{\mu}
\varepsilon_V^{\nu} \bar{u} (p_2)\gamma_5 \gamma_{\alpha} u(p_1)}
{q^2+t -m_V^2}
\frac{\mu_U^2}{Q^2+\mu_U^2}  F_{R}^A(t)
\times \nonumber \\
\times \frac{1-exp \{ -i\pi\alpha_R(t)\} }
{sin\{ \pi\alpha_R(t)\}  \Gamma \bigl (\alpha_R(t)\bigr ) }
\Bigl ( \frac{s}{Q^2} \Bigr ) ^{\alpha_R(t)-1}
\approx \nonumber \\
\approx \frac{8 C_R' \lambda_{\gamma}\lambda_{p} m_V \beta_{Rqq}\beta_{RNN}Q^2}
{Q^2-t +m_V^2}\sqrt{\frac{3 m_V \Gamma_{e^+ e^-}}{\alpha_{em}}}
\times \\
\times \frac{1-exp \{ -i\pi\alpha_R(t)\} }
{sin\{ \pi\alpha_R(t)\} \Gamma \bigl (\alpha_R(t)\bigr )}
\Bigl ( \frac{s}{Q^2} \Bigr ) ^{\alpha_R(t)}
\frac{\mu_U^2}{Q^2+\mu_U^2}F_R^A(t), \nonumber
\label{f1}
\end{eqnarray}
where $\lambda_{\gamma}$ and $\lambda_p$ are the helicities of the photon
and the proton, respectively, and $C_R^U=e_u-e_d$ for the `anomalous' $f_1$
exchange and the `normal' $f_1^\prime$ Reggeon, while $C_{a_1}^U=e_u+e_d$.
According to Eqs.~(\ref{anparr})-(\ref{parameters}),
the central values of the parameters of the exchanges are:
 $\alpha_{f_1}(t)=\alpha_{f_1}(0)=0.88$,
$\beta_{f_1 qq}\beta_{f_1 NN}=-3.04$ GeV$^{-2}$,
$\alpha_{f_1'}(t)=\alpha_{a_1}(t)=0.622+0.9\;t$,
 $\beta_{f_1' qq}\beta_{f_1' NN}=\frac{3}{5}
\beta_{A_1 qq}\beta_{A_1 NN}=13.57$ GeV$^{-2}$,
$\mu_U=1.45~\rm{GeV}$.

The axial vector form factor in the Reggeon-nucleon vertex is given
by~\cite{axialform}
\begin{equation}
F_{R}^A(t)=\frac{1}{(1-t/1.17)^2}.
\end{equation}

Only the transverse cross-section of $\rho^0$ meson electroproduction
is necessary to calculate double-spin asymmetries.
As can be seen from Fig.~\ref{rhocross},
the model described above is in fair agreement with experimental
data~\cite{rhoherm,rhoe665}.

It is, however, necessary to verify whether the assumption of $s$-channel
helicity conservation is valid for the description of
$\rho^0$ electroproduction at HERMES. Otherwise,
 a large contribution from the  spin-flip amplitude could exist
which does not enter the matrix elements related to the structure
functions and an important ingredient in the reaction mechanism
could be lost. Experimentally~\cite{hermschc} it was measured that
the violation of SCHC  is less than 10\% .

\section{The double-spin asymmetry in $\rho^0$ meson electroproduction}
Recently, HERMES has published  data on a sizeable double-spin asymmetry
in $\rho^0$ meson electroproduction on the proton~\cite{hermes}.
It is not be expected that perturbative QCD calculations can explain this result,
because for $\rho^0$ production at HERMES energy there is no hard scale
available. In general, pQCD calculations based on
two gluon-exchange in the $t$-channel predict a very small asymmetry at
$t$=0~\cite{ryskin,goloskokov}.
The phenomenological Regge approach used here takes effectively into account
non-perturbative effects of QCD which are important at HERMES energies.

The longitudinal double-spin  asymmetry $A_1^V$ for the interaction
of transverse photons with
a longitudinally polarized nucleon is defined as:
\begin{equation}
A_1^V(t)\equiv\frac
{\sigma^{1/2}_T-\sigma^{3/2}_T}
{\sigma^{1/2}_T+\sigma^{3/2}_T} =
\frac
{|M^{1/2}_T(t)|^2-|M^{3/2}_T(t)|^2}
{|M^{1/2}_T(t)|^2+|M^{3/2}_T(t)|^2}.
\end{equation}
Here $M_T^{1/2}$ and $M_T^{3/2}$ denote the transverse virtual photon
scattering amplitudes where the superscript describes the
projection of the total spin
of the photon-nucleon system to the direction of the photon momentum.
The amplitudes $M_T^{1/2,3/2}$ contain spin-independent parts made up
by exchanges with natural parity ($M_N^{1/2,3/2}$) and spin-dependent
parts made up by exchanges with unnatural parity ($M_U^{1/2,3/2}$). Between
them the following relations hold
\begin{equation}
M_N^{1/2}(t)=M_N^{3/2}(t),~~~
M_U^{1/2}(t)=-M_U^{3/2}(t).
\end{equation}
Then
\begin{eqnarray}
A_1^V(t)&=&2\frac{Re\{ M_N^{1/2}(t)\} Re\{M^{1/2}_U(t)\}}
{|M_N^{1/2}(t)|^2+|M_U^{1/2}(t)|^2} + \nonumber \\
&+&2 \frac{Im\{ M_N^{1/2}(t)\} Im\{M_U^{1/2}(t)\} }
{|M_N^{1/2}(t)|^2+|M_U^{1/2}(t)|^2}.
\label{asymm}
\end{eqnarray}

At this point this asymmetry can be analyzed qualitatively. Both types of
amplitudes contain real and
imaginary parts. The sign of the imaginary part can be determined
using the optical theorem, the sign of the real part follows from
the Regge formula given in Eq.~(\ref{prop}). The optical theorem reads
\begin{equation}
\sigma_{tot}=\frac{1}{s}Im\{ M(s,0)\}.
\end{equation}
Since the total cross-section of photon-nucleon scattering is positive,
the imaginary parts of the forward photon-nucleon scattering amplitude
for Pomeron and $f_2$ Reggeon exchanges are positive. It then follows
from Eq.~(\ref{prop}) that their real parts are negative.
In the same way we deduce that the difference of imaginary parts of
the photon-nucleon elastic scattering amplitude with anti-parallel and parallel
spins $\Delta M_U = M_U^{1/2}-M_U^{3/2}=2M_U^{1/2}$ is positive if
the contribution of the corresponding exchange
to the structure function $g_1(x,Q^2)\sim \sigma^{1/2}-
\sigma^{3/2}$ is positive. The real part of $\Delta M_U$ in this
case is also positive. If the contribution to the spin structure function
$g_1(x,Q^2)$ is negative, the imaginary and real parts of the difference
$\Delta M_U$ are negative.

It is worthwhile to note that a direct estimate
of the vector meson production asymmetry based on the relation $A_1^V\approx
2A_1^{DIS}$~\cite{nikolaev}, even in the context of SCHC, does inherently not
include the real part of the vector meson production amplitude as the DIS
data on $g_1$ and $F_2$ are
connected only to the imaginary part. For a full description, however, a way
must be found to construct the real part which, in this paper, was chosen to be
the Regge approach described above.

For the `anomalous' $f_1$ exchange, the real part is much larger than the
imaginary part and therefore this exchange should give a positive
contribution to the asymmetry.
For $f_1'$ and $a_1$ exchanges, real and imaginary parts of the amplitudes
contribute with different signs to the asymmetry. For scattering on
the proton, their real parts give a negative asymmetry, and the imaginary
parts give a positive one. This is also the case for $f_1'$ exchange when
scattering on the neutron and on the deuteron. For $a_1$ exchange in the case
of scattering on the neutron, the real part of the amplitude leads to a
positive double-spin asymmetry and the imaginary part leads to a negative one.
Altogether, this discussion applies only to the region of small $|t|$ where
the corresponding amplitudes do not change sign ($\alpha=0$).
Since the elastic cross-section originates mainly from the small $|t|$ region,
we expect this qualitative analysis to be valid.

  The calculated predictions for double-spin asymmetries are shown in
       Figs. 6-13. Shaded areas were chosen to illustrate the range
       in the predictions obtained when different values for the `anomalous'
       $f_1$ intercept are used. The upper limit of
this range,  $ 0.74 - 0.93$
       (cf. Eq.~\ref{anparr}),
is determined by the restriction that it is necessary to
       reproduce the transverse cross section shown in Fig. 5. Intercept
       values very close to unity are excluded, because otherwise the
       propagator in Eq. (1) would yield a too large contribution for
       $\alpha_{f_1}\rightarrow 1$.

From the figures
 it is evident that the contributions of the secondary
$f_1'$ and $a_1$ Reggeons to the asymmetry are not small.
 At HERMES energies, they are of about the same size as
 the contribution of the `anomalous' $f_1$ exchange which we
considered previously~\cite{our2} within another approach.

More information about the flavor composition of the asymmetry can be
obtained when considering  the case of the  deuteron and the neutron. For the
deuteron the isovector exchange contribution to $A_1^V$ vanishes, and for
the neutron it comes with the opposite sign and partly cancels the
regular isoscalar part. By this reason the `anomalous' part can be observed
best when studying data on the neutron.

 The dependencies of the double-spin asymmetry $A^1_V$ on various
                kinematic variables are displayed in these figures for proton
                and deuteron. The Regge-based predictions were calculated for
                the HERMES kinematics
, $\langle W\rangle$=4.9 GeV, $\langle
Q^2\rangle$=1.7 GeV$^2$, $\langle x\rangle$=0.07,
 and compared  to measurements of $\rho^0$ electroproduction at
HERMES~\cite{hermes,hermesd,hermesd1} on proton and deuteron.
Although qualitative agreement can be seen only in the case of the proton,
it is evident that experimental data with considerably improved precision
are required.

\section{Conclusions}
In summary, the nucleon structure functions $F_2(x,Q^2)$ and $g_1(x,Q^2)$
were analyzed in the framework of a Regge approach. From the data on deep
inelastic scattering on the proton, neutron and deuteron we derived the
parameters of Reggeons with
natural and unnatural parities in the region $Q^2$=1$\div$3 GeV$^2$.
Using this parameterization and a non-relativistic  model
of $\rho^0$ meson formation which provides a fair description of
the cross-section, we calculated the double-spin asymmetry
of $\rho^0$ meson electroproduction at HERMES energies.

In this study we have used a unified  approach to both  DIS
spin-dependent structure functions and
vector meson electroproduction, in the context of $s$-channel helicity
conservation. In this approach the obtained large value of
the double-spin asymmetry in $\rho^0$ meson production is correlated with the
anomalous behavior of the flavor-singlet part of the
structure function $g_1(x,Q^2)$ at small $x$.
In the case that future measurements will not confirm such a large asymmetry,
for a Regge-type analysis it would have to be concluded
that the `anomalous' $f_1$ exchange is not a  simple Regge pole
but has a more complicated structure, e.g. a Regge cut.

\section*{Acknowledgments}
We are grateful to N.Bianchi, S.V.Goloskokov and O.V.Teryaev for useful
discussions.
This work was supported by the following grants RFBR-01-02-16431,
INTAS-2000-366, EC-IHP-HPRN-CT-2000-00130, MCyT-BFM2001-0262, GV01-216,
and  by the Heisenberg-Landau program.
A.V. thanks ICTP for the warm hospitality extended to him
during his stay when a part of this work was completed and
V.V. acknowledges the hospitality of Seoul National University during
the last stages of this work.


\begin{figure*}
\centering
\epsfig{file=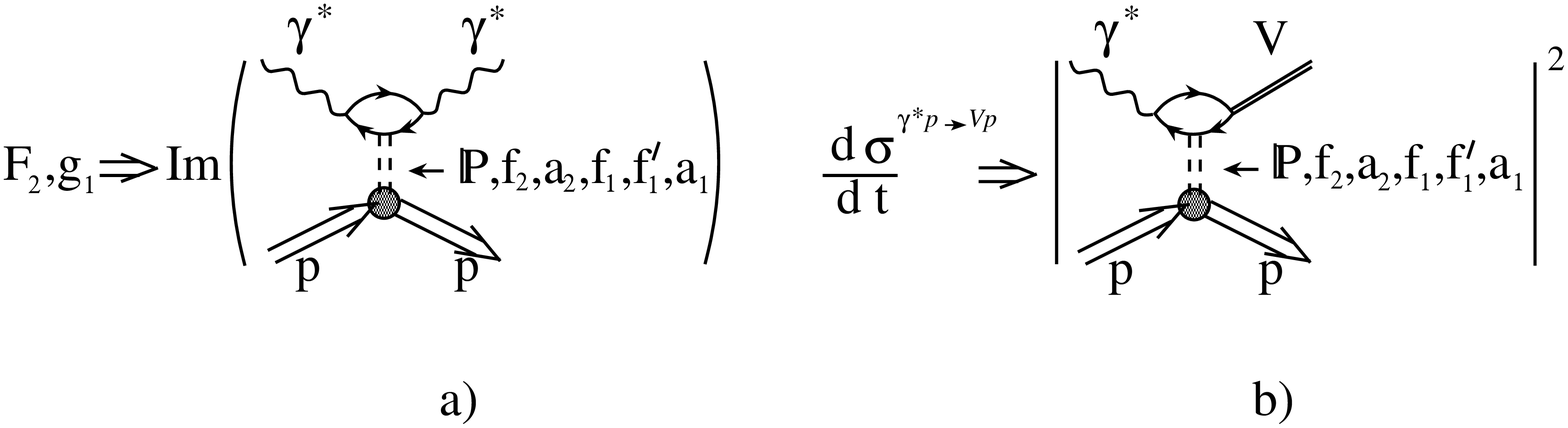,width=0.9\hsize}
\caption{S- channel helicity conservation implies that the same
exchanges describe the structure functions ($F_2$ and $g_1$)  of inclusive deep
inelastic lepton scattering [panel a)] and of vector meson production
processes at high energy [panel b)].}
\end{figure*}
\begin{figure*}
\centering
\begin{minipage}[c]{0.45\hsize}
\centering
\epsfig{file=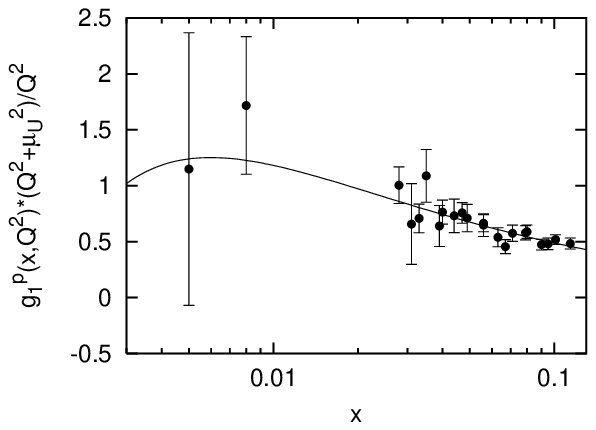,width=0.95\hsize}
\caption{The result of the fit of the proton
structure function $g_1(x,Q^2)$ in the Regge region
($x < $0.1) and at Q$^2$ = 1~-~3~GeV$^2$ in comparison with
the data \cite{g1dat}.}
\label{g1pp}
\end{minipage}
\hspace*{1cm}
\begin{minipage}[c]{0.45\hsize}
\centering
\epsfig{file=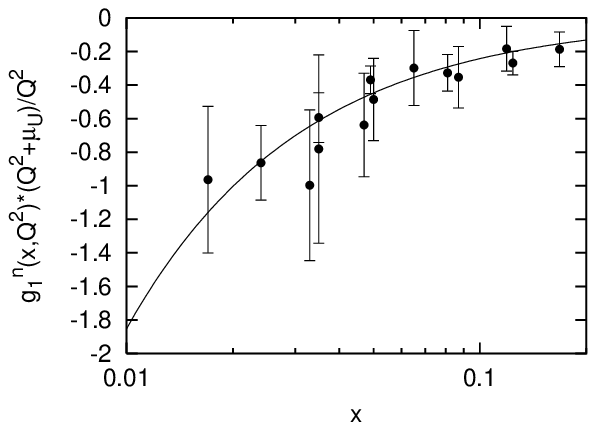,width=0.95\hsize}
\caption{The result of the fit of the neutron
structure function $g_1(x,Q^2)$ in the Regge region
($x < $0.1) and at Q$^2$ = 1~-~3~GeV$^2$ in comparison with
the data \cite{g1dat}.}
\label{g1nn}
\end{minipage}
\end{figure*}
\begin{figure*}
\centering
\begin{minipage}[c]{0.45\hsize}
\centering
\epsfig{file=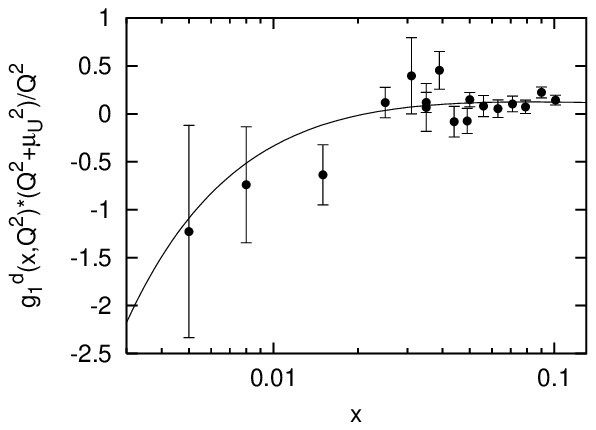,width=0.95\hsize}
\caption{The result of the fit of the deuteron
structure function $g_1(x,Q^2)$ in the Regge region
($x < $0.1) and at Q$^2$ = 1~-~3~GeV$^2$ in comparison with
the data \cite{g1dat}.}
\label{g1dd}
\end{minipage}
\hspace*{1cm}
\begin{minipage}[c]{0.45\hsize}
\centering
\epsfig{file=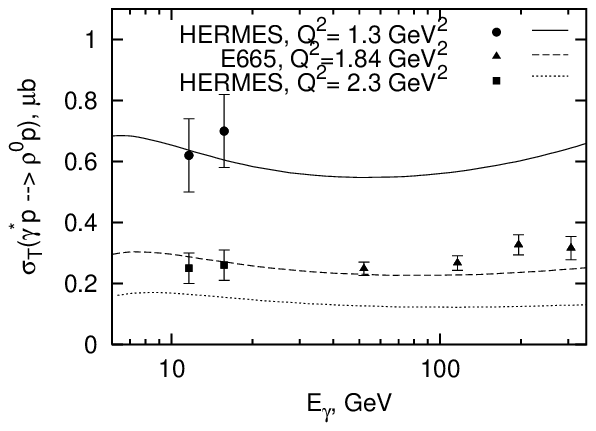,width=0.95\hsize}
\caption{The cross-section of $\rho^0$ electroproduction by transverse photons
is shown.
The data points  have been obtained from the
published total and longitudinal cross-sections measured at
HERMES~\cite{rhoherm} and E665~\cite{rhoe665}.
The solid, dashed and dotted lines represent the model calculations
at the given $Q^2$ values.}
\label{rhocross}
\end{minipage}
\end{figure*}

\begin{figure*}
\centering
\begin{minipage}[c]{0.45\hsize}
\centering
\epsfig{file=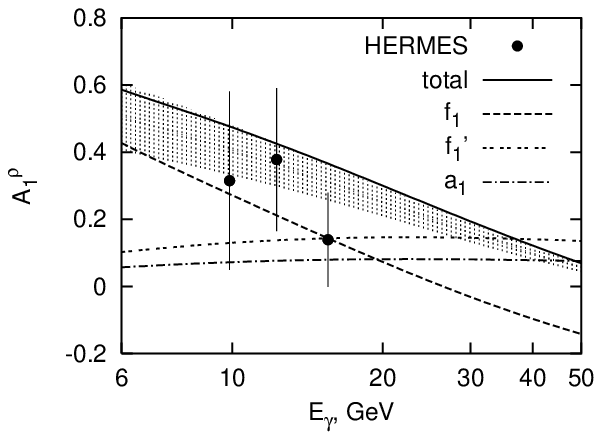,width=\hsize}
\caption{The $E_{\gamma^*}$-dependence of the double-spin asymmetry $A_1^\rho$
on the proton compared to data calculated from results
of HERMES \cite{hermes}. The shaded area corresponds to the interval
of allowed values $0.72\div 0.93$ for  the  anomalous $f_1$ intercept
(see section V).}
\label{nu1}
\end{minipage}
\hspace*{1cm}
\begin{minipage}[c]{0.45\hsize}
\centering
\epsfig{file=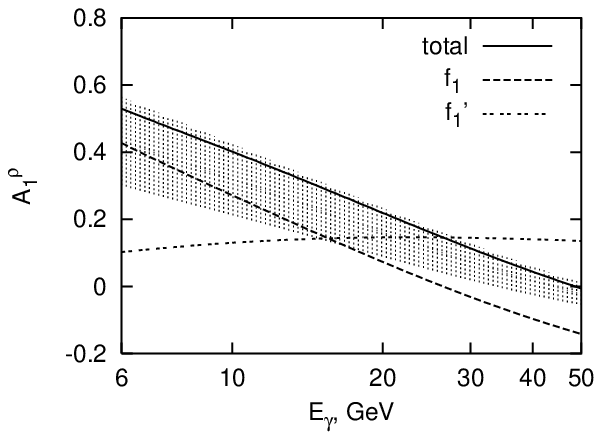,width=\hsize}
\caption{The $E_{\gamma^*}$-dependence of the double-spin asymmetry $A_1^\rho$
on the deuteron is shown. The notations are the same as in Fig.~\ref{nu1}.}
\label{nu2}
\end{minipage}
\end{figure*}

\begin{figure*}
\centering
\begin{minipage}[c]{0.45\hsize}
\centering
\epsfig{file=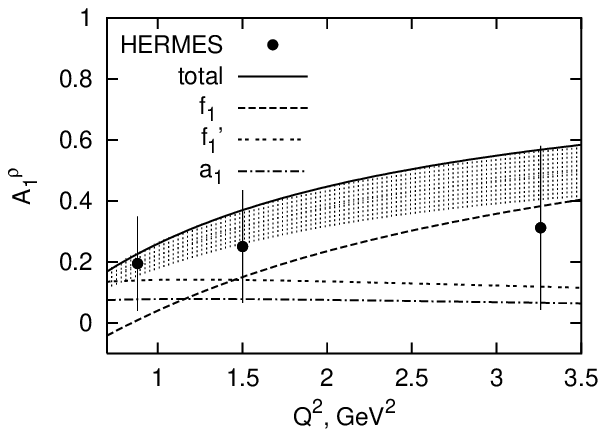,width=\hsize}
\caption{The $Q^2$-dependence of the double-spin asymmetry $A_1^\rho$ on the
proton, compared to results of HERMES \cite{hermes}.
The notations are the same as in Fig.~\ref{nu1}.}
\label{q1}
\end{minipage}
\hspace*{1.cm}
\begin{minipage}[c]{0.45\hsize}
\centering
\epsfig{file=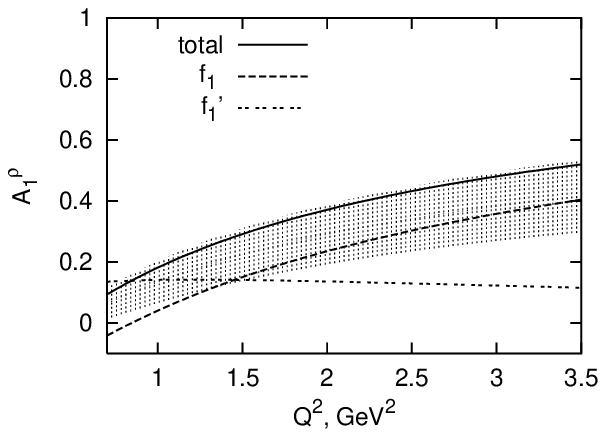,width=\hsize}
\caption{The $Q^2$-dependence of the double-spin asymmetry $A_1^\rho$ on the
deuteron is shown. The notations are the same as in Fig.~\ref{nu1}.}
\label{q2}
\end{minipage}
\end{figure*}

\begin{figure*}
\centering
\begin{minipage}[c]{0.45\hsize}
\centering
\epsfig{file=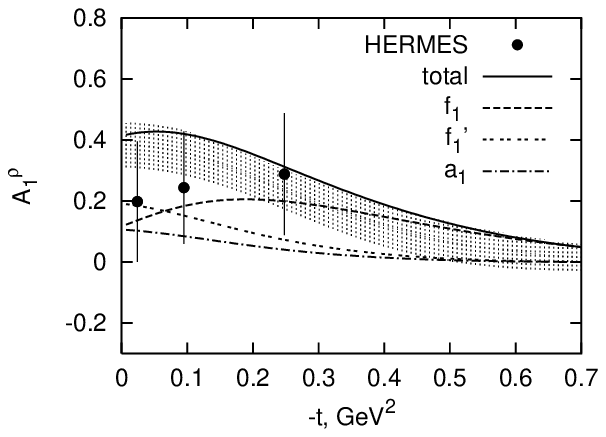,width=\hsize}
\caption{The $-t$-dependence of the double-spin asymmetry $A_1^\rho$ on the
proton, compared to results of HERMES \cite{hermes}.
The notations are the same as in Fig.~\ref{nu1}.}
\label{t1}
\end{minipage}
\hspace*{1cm}
\begin{minipage}[c]{7.7cm}
\centering
\epsfig{file=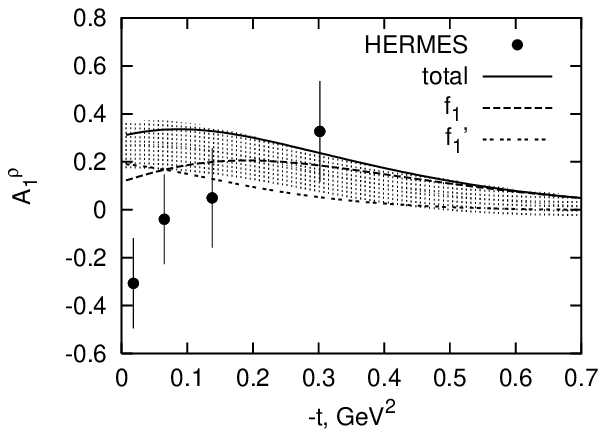,width=\hsize}
\caption{The $-t$-dependence of the double-spin asymmetry $A_1^\rho$ on the
deuteron, compared to preliminary results of HERMES \cite{hermesd1}.
The notations are the same as in Fig.~\ref{nu1}.}
\label{t2}
\end{minipage}
\end{figure*}

\begin{figure*}
\centering
\begin{minipage}[c]{0.45\hsize}
\centering
\epsfig{file=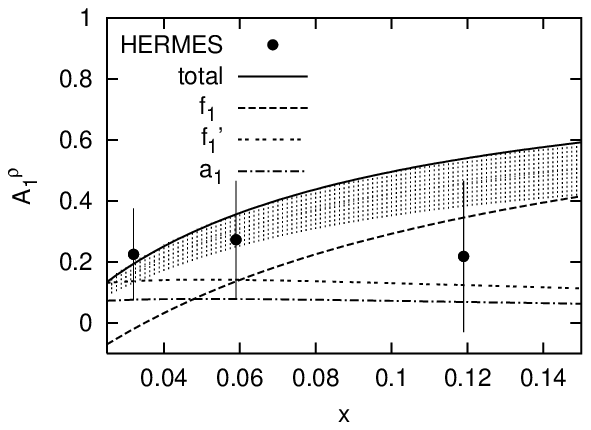,width=\hsize}
\caption{The $x$-dependence of the double-spin asymmetry $A_1^\rho$ on
the proton, compared to results of HERMES \cite{hermes}.
The notations are the same as in Fig.~\ref{nu1}.}
\label{x1}
\end{minipage}
\hspace*{1cm}
\begin{minipage}[c]{0.45\hsize}
\centering
\epsfig{file=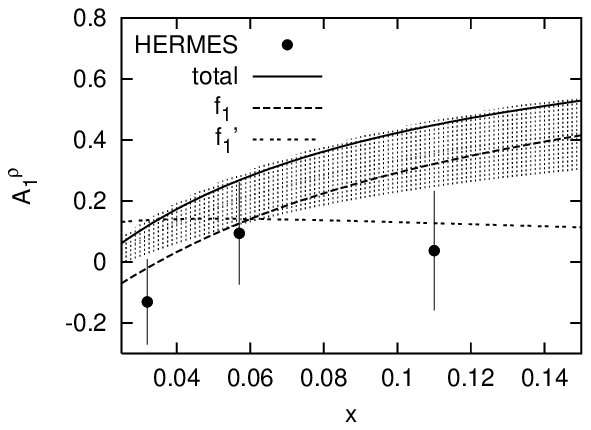,width=\hsize}
\caption{The $x$-dependence of the double-spin asymmetry $A_1^\rho$ on
the deuteron, compared to preliminary results of HERMES \cite{hermesd}.
The notations are the same as in Fig.~\ref{nu1}.}
\label{x2}
\end{minipage}
\end{figure*}


\begin{thebibliography}{99}

\bibitem{g1dat}
B.~Adeva {\it et al.}  [Spin Muon Collaboration],
Phys.\ Rev.\ D {\bf 58}, 112001 (1998).\\
K.~Abe {\it et al.}  [E143 collaboration],
Phys.\ Rev.\ D {\bf 58}, 112003 (1998). \\
A.~Airapetian {\it et al.}  [HERMES Collaboration],
Phys.\ Lett.\ B {\bf 442}, 484 (1998). \\
P.~L.~Anthony {\it et al.}  [E142 Collaboration],
Phys.\ Rev.\ D {\bf 54}, 6620 (1996). \\
K.~Abe {\it et al.}  [E154 Collaboration],
Phys.\ Rev.\ Lett.\  {\bf 79}, 26 (1997). \\
K.~Ackerstaff {\it et al.}  [HERMES Collaboration],
Phys.\ Lett.\ B {\bf 404}, 383 (1997).

\bibitem{rysking1}
J.~Bartels, B.~I.~Ermolaev and M.~G.~Ryskin,
Z.\ Phys.\ C {\bf 72}, 627 (1996).

\bibitem{bfkl}
E.~A.~Kuraev, L.~N.~Lipatov and V.~S.~Fadin,
Sov.\ Phys.\ JETP {\bf 45}, 199 (1977)
[Zh.\ Eksp.\ Teor.\ Fiz.\  {\bf 72}, 377 (1977)].

\bibitem{lipatov}
V.~S.~Fadin and L.~N.~Lipatov,
Phys.\ Lett.\ B {\bf 429}, 127 (1998).

\bibitem{our1}
N.~I.~Kochelev {\it et al.},
Phys.\ Rev.\ D {\bf 61}, 094008 (2000).

\bibitem{our11} N.~I.~Kochelev {\it et al.},
Nucl.\ Phys.\ Proc.\ Suppl.\  {\bf 99A}, 24 (2001).

\bibitem{our2}
N.~I.~Kochelev, D.~P.~Min, V.~Vento and A.~V.~Vinnikov,
Phys.\ Rev.\ D {\bf 65}, 097504 (2002).

\bibitem{our3}
Y.~Oh, N.~I.~Kochelev, D.~P.~Min, V.~Vento and A.~V.~Vinnikov,
Phys.\ Rev.\ D {\bf 62}, 017504 (2000).

\bibitem{collins}
P.~D.~Collins,
``An Introduction To Regge Theory And High-Energy Physics,''
{\it  Cambridge 1977, 445p}.

\bibitem{levin}
E.~Levin, TAUP-2465-97, DESY-97-213,
arXiv:hep-ph/9710546.

\bibitem{simonov}
A.~B.~Kaidalov and Y.~A.~Simonov,
Phys.\ Lett.\ B {\bf 477}, 163 (2000).

\bibitem{kharzeev}
D.~Kharzeev and E.~Levin,
Nucl.\ Phys.\ B {\bf 578}, 351 (2000).

\bibitem{shuryak}
E.~V.~Shuryak and I.~Zahed,
Phys.\ Rev.\ D {\bf 62}, 085014 (2000).

\bibitem{laget97}
M.~Guidal, J.~M.~Laget and M.~Vanderhaeghen,
Nucl.\ Phys.\ A {\bf 627}, 645 (1997).

\bibitem{manaenk}
S.~I.~Manaenkov,
arXiv:hep-ph/9903405.

\bibitem{landtot}
A.~Donnachie and P.~V.~Landshoff,
Phys.\ Lett.\ B {\bf 296}, 227 (1992).

\bibitem{pdg}
D.~E.~Groom {\it et al.}  [Particle Data Group Collaboration],
Eur.\ Phys.\ J.\ C {\bf 15}, 1 (2000).

\bibitem{landnacht}
P.~V.~Landshoff and O.~Nachtmann,
Z.\ Phys.\ C {\bf 35}, 405 (1987).

\bibitem{kaidalov}
A.~Capella, A.~Kaidalov, C.~Merino and J.~Tran Thanh Van,
Phys.\ Lett.\ B {\bf 343}, 403 (1995).

\bibitem{H1}
C.~Adloff {\it et al.}  [H1 Collaboration],
Phys.\ Lett.\ B {\bf 520}, 183 (2001).

\bibitem{ZEUS}
J.~Breitweg {\it et al.}  [ZEUS Collaboration],
Eur.\ Phys.\ J.\ C {\bf 7}, 609 (1999).

\bibitem{allm}
H.~Abramowicz, E.~M.~Levin, A.~Levy and U.~Maor,
Phys.\ Lett.\ B {\bf 269}, 465 (1991). \\
A.~Donnachie and P.~V.~Landshoff,
Z.\ Phys.\ C {\bf 61}, 139 (1994).

\bibitem{f2data}
M.~R.~Adams {\it et al.}  [E665 Collaboration],
Phys.\ Rev.\ D {\bf 54}, 3006 (1996).\\
M.~Arneodo {\it et al.}  [New Muon Collaboration],
Nucl.\ Phys.\ B {\bf 483}, 3 (1997).\\
D.~Allasia {\it et al.},
Z.\ Phys.\ C {\bf 28}, 321 (1985).\\
S.~Aid {\it et al.}  [H1 Collaboration],
Nucl.\ Phys.\ B {\bf 470}, 3 (1996).\\
C.~Adloff {\it et al.}  [H1 Collaboration],
Nucl.\ Phys.\ B {\bf 497}, 3 (1997).\\
M.~Derrick {\it et al.}  [ZEUS Collaboration],
Z.\ Phys.\ C {\bf 69}, 607 (1996).\\
J.~Breitweg {\it et al.}  [ZEUS Collaboration],
Eur.\ Phys.\ J.\ C {\bf 7}, 609 (1999).

\bibitem{teryaev}
J.~Soffer and O.~V.~Teryaev,
Phys.\ Rev.\ D {\bf 56}, 1549 (1997).\\
S.~D.~Bass and M.~M.~Brisudova,
Eur.\ Phys.\ J.\ A {\bf 4}, 251 (1999).

\bibitem{bianchi}
N.~Bianchi and E.~Thomas,
Phys.\ Lett.\ B {\bf 450}, 439 (1999).

\bibitem{hermschc}
M.~Tytgat,
DESY-THESIS-2001-018.

\bibitem{rosen}
L.~Rosenberg,
Phys.\ Rev.\  {\bf 129}, 2786 (1963).

\bibitem{efremov}
M.~Anselmino, A.~Efremov and E.~Leader,
Phys.\ Rept.\  {\bf 261}, 1 (1995)
[Erratum-ibid.\  {\bf 281}, 399 (1997)].

\bibitem{landrho}
A.~Donnachie and P.~V.~Landshoff,
Phys.\ Lett.\ B {\bf 185}, 403 (1987).

\bibitem{laget}
J.~M.~Laget,
Phys.\ Lett.\ B {\bf 489}, 313 (2000).

\bibitem{cudell}
I.~Royen and J.~R.~Cudell,
Nucl.\ Phys.\ B {\bf 545}, 505 (1999).

\bibitem{axialform}
A.~Liesenfeld {\it et al.}  [A1 Collaboration],
Phys.\ Lett.\ B {\bf 468}, 20 (1999).

\bibitem{rhoherm}
A.~Airapetian {\it et al.}  [HERMES Collaboration],
Eur.\ Phys.\ J.\ C {\bf 17}, 389 (2000).

\bibitem{rhoe665}
M.~R.~Adams {\it et al.}  [E665 Collaboration],
Z.\ Phys.\ C {\bf 74}, 237 (1997).

\bibitem{hermes}
A.~Airapetian {\it et al.}  [HERMES Collaboration],
Phys.\ Lett.\ B {\bf 513}, 301 (2001).

\bibitem{ryskin}
M.~Vanttinen and L.~Mankiewicz,
Phys.\ Lett.\ B {\bf 440}, 157 (1998),\\
M.~Vanttinen and L.~Mankiewicz,
Phys.\ Lett.\ B {\bf 434}, 141 (1998).

\bibitem{goloskokov}
S.~V.~Goloskokov,
Eur.\ Phys.\ J.\ C {\bf 11}, 309 (1999).

\bibitem{nikolaev}
H.~Fraas,
Nucl.\ Phys.\ B {\bf 113}, 532 (1976).\\
N.~N.~Nikolaev,
Nucl.\ Phys.\ Proc.\ Suppl.\  {\bf 79}, 343 (1999).

\bibitem{hermesd}
K. Lipka ({ \it for the HERMES Collaboration}), 
World Scientific, Proc. of PHOTON 2001, 186 (2002).

\bibitem{hermesd1}
K. Lipka, talk at An International Conference on The Structure and
Interactions of the Photon (PHOTON 2001), Ascona Switzerland September
2nd-7th 2001,
http://hep-proj-photon2001.web.cern.ch/hep-proj-photon2001/proc/proc.htm

\end{thebibliography}
\end{document}